\documentclass[10pt,conference]{IEEEtran} 
\IEEEoverridecommandlockouts

\usepackage{cite}
\usepackage{amsmath,amssymb,amsfonts}
\usepackage{graphicx}
\usepackage{textcomp}
\usepackage{xcolor}
\def\BibTeX{{\rm B\kern-.05em{\sc i\kern-.025em b}\kern-.08em
    T\kern-.1667em\lower.7ex\hbox{E}\kern-.125emX}}
\usepackage{booktabs}
\usepackage{xcolor}
\usepackage[most]{tcolorbox}
\usepackage{tabularx}

\usepackage{soul}
\usepackage{multirow}
\usepackage{listings}
\usepackage{fancybox}
\usepackage{enumitem}
\usepackage{graphicx}
\usepackage{color}
\usepackage{array}
\usepackage{amsmath}
\usepackage{xspace}
\usepackage{url}
\usepackage{tikz}
\usepackage{caption}
\usepackage{pgfplots}
\usepackage{balance}
\usepackage{subcaption}
\pgfplotsset{compat=1.16}
\usepackage{pgf-pie}
\usetikzlibrary{pgfplots.statistics,calc}
\usepackage{algpseudocode}
\usepackage{algorithm}
\usepackage{adjustbox}
\usepackage{makecell}
\usepackage{listings}
\usepackage{svg}
\usepackage{listings}
\usepackage{xcolor}

\usepackage{tikz}

\newcommand*\circled[1]{\tikz[baseline=(char.base)]{
            \node[shape=circle,draw,inner sep=2pt] (char) {#1};}}

\usepackage{tcolorbox}
\makeatletter
\newcommand{\mybox}[1]{%
	\setbox0=\hbox{#1}%
	\setlength{\@tempdima}{\dimexpr\wd0+13pt}%
	\begin{tcolorbox}[boxrule=0.5pt, colback=white, arc=4pt,
		left=6pt,right=6pt,top=6pt,bottom=6pt,boxsep=0pt]
		#1
	\end{tcolorbox}
}

\definecolor{songcolor}{RGB}{191,191,191}

\newcommand{\tool}{\textit{A2HTrans}}

\definecolor{codegreen}{rgb}{0,0.6,0}
\definecolor{codegray}{rgb}{0.5,0.5,0.5}
\definecolor{codepurple}{rgb}{0.58,0,0.82}
\definecolor{backcolour}{rgb}{0.95,0.95,0.92}
\lstdefinestyle{Cstyle}{
    language=Java,
    aboveskip=3mm,
    showstringspaces=false,
    columns=flexible,
    numbers=left,
    backgroundcolor=\color{backcolour},
    commentstyle=\color{codegreen},
    keywordstyle=\color{magenta},
    numberstyle=\tiny\color{codegray},
    stringstyle=\color{codepurple},
    basicstyle=\scriptsize\ttfamily,
    breakatwhitespace=false,         
    breaklines=true,                 
    captionpos=b,                    
    keepspaces=false,                 
    numbersep=5pt,                  
    showspaces=false,                
    showstringspaces=false,
    showtabs=false,                  
    tabsize=2,
    escapeinside=``,
}

\begin{document}

\title{Beyond Code Translation: Agentic Android-to-HarmonyOS Migration}

\author{
\IEEEauthorblockN{
Yinghang Ma\IEEEauthorrefmark{1},
Wuyang Dai\IEEEauthorrefmark{1},
Song Wang\IEEEauthorrefmark{1},
Filipe Cogo\IEEEauthorrefmark{2},
Zhen Ming (Jack) Jiang\IEEEauthorrefmark{1}
\thanks{Yinghang Ma and Wuyang Dai contributed equally to this work.}
}
\IEEEauthorblockA{
\IEEEauthorrefmark{1}York University, Toronto, Canada\\
\{coppelia,ddai2002,wangsong,zmjiang\}@yorku.ca
}
\IEEEauthorblockA{
\IEEEauthorrefmark{2}Queen's University, Kingston, Canada\\
filipe.cogo@queensu.ca
}
}

\maketitle
\begin{abstract} 
Migrating Android applications to HarmonyOS is an important yet challenging software engineering task. Existing large language model (LLM)-based migration approaches primarily translate code at the method or file level, overlooking project-wide dependencies and workflow coordination. Moreover, ArkTS, the primary language for HarmonyOS, is relatively low-resource, and migrated applications often lack sufficient tests for reliable verification, making end-to-end migration error-prone. We present {\tool}, an agentic framework for end-to-end Android-to-HarmonyOS application migration. Rather than performing isolated code translation, {\tool} decomposes applications into functionality-aware migration tasks, prioritizes them through dependency analysis, and orchestrates translation, validation, and iterative repair within a unified execution workflow. To improve migration reliability, {\tool} employs a tool-augmented verification strategy that combines compiler-based syntax checking with LLM-based semantic validation and automated repair. We evaluate {\tool} on 14 Android applications spanning diverse sizes and domains using Claude-Sonnet-4.6 as the underlying migration model. Compared with Claude Code, {\tool} preserves 8.3\% more functional completeness, improves behavioral correctness by 8.0\%, and reduces migration time by 28.8\%. These results demonstrate that structured agentic execution substantially improves the reliability and efficiency of project-level cross-platform software migration.
\end{abstract}

\section{Introduction}
\label{sec:intro}

Mobile software ecosystems evolve rapidly, frequently requiring developers to migrate existing applications across platforms~\cite{ribeiro2012survey}. Since 2021, Huawei has begun manufacturing HarmonyOS-based phones instead of Android-based phones. To ensure a vibrant application ecosystem for users, it is therefore vital to migrate existing Android apps to HarmonyOS\footnote{\url{https://appinchina.co/the-complete-guide-to-harmonyos-next/}}. 
Although Android and HarmonyOS share certain architectural concepts, they differ substantially in programming languages (i.e., Java/Kotlin vs. ArkTS), UI frameworks, permission models, application lifecycle management, APIs, and other features. 
Currently, migrating real-world Android applications to HarmonyOS is a challenging and labor-intensive engineering task, as developers must rely on a combination of manual interpretation of platform documentation, extensive code modifications, and only partially automated migration tools. 

Existing LLM-based code translation (e.g., TRANSAGENT~\cite{liang2023transagent}, AlphaTrans~\cite{ibrahimzada2025alphatrans}, and Repoagent~\cite{luo2024repoagent}) primarily focuses on method-level code-to-code transformations, 
leaving higher-level application logic largely unaddressed. Building on this, Zeng et al.~\cite{zeng2025pilot} recently analyzed translation bugs in LLM-based automatic migration from Android to iOS, highlighting the challenges of cross-platform application translation but focusing mainly on syntax errors in the translated code. In contrast, end-to-end translation of real-world mobile applications introduces additional challenges, including cross-platform semantic mismatches, UI layout differences, resource-handling conventions, and platform-specific configurations, making it a substantially more complex problem that remains largely unexplored.

Motivated by these challenges, in this paper, we report on our experience building an LLM-based framework, called {\tool}, that combines an agentic translation pipeline with model adaptation via fine-tuning, enabling end-to-end cross-platform migration of real-world mobile applications. 
Given an Android application, {\tool} adopts an agentic translation workflow that iteratively performs planning, action, and verification across multiple stages. First, the framework analyzes the application structure to identify components and their dependencies, and plans the migration by decomposing the codebase into feature-oriented clusters of related components, classes, and methods. Based on this plan, {\tool} uses a general-purpose LLM (e.g., Claude-Sonnet-4.5) to infer high-level application features from each cluster and to guide feature-aware translation. 
The agent then executes translation actions and generates a valid ArkTS project scaffold and configuration. During this process, {\tool} invokes the ArkTS compiler to detect syntax errors and iteratively corrects ArkTS-specific issues using a fine-tuned, domain-specific lightweight LLM (e.g., DeepSeek-16B).   
Finally, {\tool} leverages the general-purpose LLM to verify the translated features by inspecting their implementations and checking alignment between the original Android application and the translated HarmonyOS application. When misalignments are detected, {\tool} iteratively refines the translation. 
To fine-tune a domain-specific lightweight LLM, we construct a dataset comprising 1.7K ArkTS specification–code pairs derived from official HarmonyOS examples. This dataset enables the LLM to understand ArkTS-specific syntax, APIs, and coding conventions, enabling it to automatically detect and correct ArkTS-specific syntax errors during the translation process. 

We evaluated {\tool} on 14 real-world Android applications of varying size and complexity. To assess its effectiveness, we conducted a manual UI/behavior review using five 0–100 metrics: functional completeness, functional correctness, functional appropriateness, operability, and UI alignment with the original Android application. Scores were assigned by two reviewers and averaged to mitigate individual bias. In addition, we report translation time as a proxy for effort and use pre-repair build-error counts to characterize syntactic deployability prior to repair. We compare {\tool} against two baselines: a basic LLM-based prompting approach and the Claude Code CLI~\cite{claude}.

The results show that {\tool} improves migration quality while reducing translation time. Compared with Claude Code, {\tool} increases functional completeness by 8.3\%, functional correctness by 8.0\%, functional appropriateness by 3.0\%, UI operability by 5.3\%, and UI alignment by 5.5\%. It also reduces average translation time by 28.8\%. We do not use token count as a primary efficiency metric because Claude Code and {\tool} rely on different context-management and caching behaviors, making raw token counts not directly comparable. Overall, these results indicate that {\tool} better preserves task-level functionality and behavior while requiring less translation time. 

This paper makes the following contributions:

\begin{itemize}

\item We address ArkTS’s low-resource challenge by curating training data from official documentation, in the form of 1.7K ArkTS specification–code pairs. We demonstrate the usage of this data by fine-tuning LLMs to correct compiler-detected ArkTS syntax and API errors within our agentic migration workflow.

\item We propose {\tool}, the first agentic framework for end-to-end migration of Android applications to HarmonyOS. We also demonstrate the superior performance of our approach, which performs feature-guided, dependency-aware translation across components and project configurations rather than method-level code-to-code translations. 

\item We design a tool-augmented validation strategy for {\tool}, i.e., compiler-based syntax checking with LLM-based semantic validation, showing that it significantly improves the correctness of translated ArkTS applications.

\item We conduct an empirical evaluation on 14 real-world Android projects. The results show that {\tool} outperforms the Claude Code CLI baseline, achieving better preservation of application functionality, task-level behavior, and interface consistency in Android-to-HarmonyOS migration.

\end{itemize}

The rest of this paper is organized as follows: Section~\ref{sec:background} introduces the background of this work, Section~\ref{sec:3} presents our multi-agent pipeline for Android-to-HarmonyOS migration, Section~\ref{sec:4} shows the settings of our case study, Section~\ref{sec:result} presents the results, 
Section~\ref{sec:lesson} discusses the lessons learned from translation issues, 
Section~\ref{sec:threats} presents the threats to validity of this work, and Section~\ref{sec:conclusion} concludes this work. 

\section{Background and Related Work}
\label{sec:background}

\subsection{HarmonyOS and ArkTS}
\label{sec:2.1}

HarmonyOS is a distributed operating system designed to run across smartphones, wearables, smart home devices, and in-vehicle platforms, supporting seamless cross-device collaboration~\cite{li2025software}. Application development on HarmonyOS primarily relies on ArkTS, the platform’s official programming language. ArkTS builds upon TypeScript while introducing platform-specific extensions and language features specifically designed for cross-device and multi-platform application development~\cite{aytekin2026automating}. 
A notable example is ArkUI, a declarative user interface framework that enables developers to define UI components through reactive data bindings, thereby simplifying interface construction and state management.  
Despite these benefits, ArkTS is a recently introduced language, with very limited publicly available data (e.g., code examples, documentation, and tutorials), posing challenges for developers and limiting the effectiveness of existing LLMs for code generation and translation.  
Recently, Wang et al.~\cite{wang2024a2h} proposed leveraging LLMs to automatically translate Android XML layouts into HarmonyOS ArkUI code, showing promising results for UI-level migration. However, their approach addresses only isolated aspects of migration, and UI translation alone is insufficient for real-world applications, where overall correctness depends on multiple platform-specific behaviors. 

\subsection{Code Translation}
\label{sec:2.2}

Code translation aims to convert source code from one programming language to another while preserving the original program’s functionality~\cite{xia2023empirical}. Traditional code translation approaches~\cite{C2Rust, Sharpen, phan2017statistical, pawlak2016spoon} primarily rely on rule-based techniques and syntax-directed transformation schemes~\cite{hong2023improving}. While effective in limited settings, these methods often struggle to scale to complex software systems or to capture higher-level program semantics.

More recently, LLMs have demonstrated strong capabilities in code translation. However, prior studies show that LLMs may introduce subtle semantic bugs due to incomplete contextual understanding or misinterpretation of language-specific constructs~\cite{pan2024lost}. These findings highlight the need for more advanced translation approaches that operate at the project level, incorporating contextual and structural information to reduce translation errors.
Jana et al.~\cite{jana2024cotran} propose fine-tuning LLMs using compiler and symbolic-execution-based feedback to support translation across multiple programming languages,  
demonstrating the potential of feedback-driven LLM adaptation for improving translation correctness. 
Bairi et al.~\cite{bairi2024codeplan} introduce 
CodePlan, a task-agnostic framework for inter-repository coding tasks.   
Ibrahimzada et al.~\cite{ibrahimzada2024alphatrans} proposed \textit{AlphaTrans}, a neuro-symbolic approach to automate repository-level code translation. 
Additionally, systems such as RepoAgent~\cite{luo2024repoagent}, \textit{CodePlan}~\cite{jain2023codeplan}, and \textit{TRANSAGENT}~\cite{liang2023transagent} 
explore repository-level code translation by integrating planning and multi-agent strategies to manage the complexities of larger codebases. These approaches, however, generally target mainstream programming languages such as Rust, C, and C++, and few have been applied to low-resource languages like ArkTS. 
Recently, Zeng et al.~\cite{zeng2025pilot} manually analyzed the translation bugs in LLM-based automatic migration from Android to iOS, highlighting the challenges of cross-platform application translation for low-resource languages.

Despite these advancements, the effectiveness and applicability of LLM-based approaches for mobile application migration, particularly across different platforms such as Android and HarmonyOS, remain largely unexplored. In this work, we investigate the capabilities and limitations of LLM-based agentic migration 
for mobile applications, using Android-to-HarmonyOS migration as a representative and practically significant case study.

\section{Approach}
\label{sec:3}



\begin{figure}[t!]
  \centering
  \includegraphics[width=0.48\textwidth]{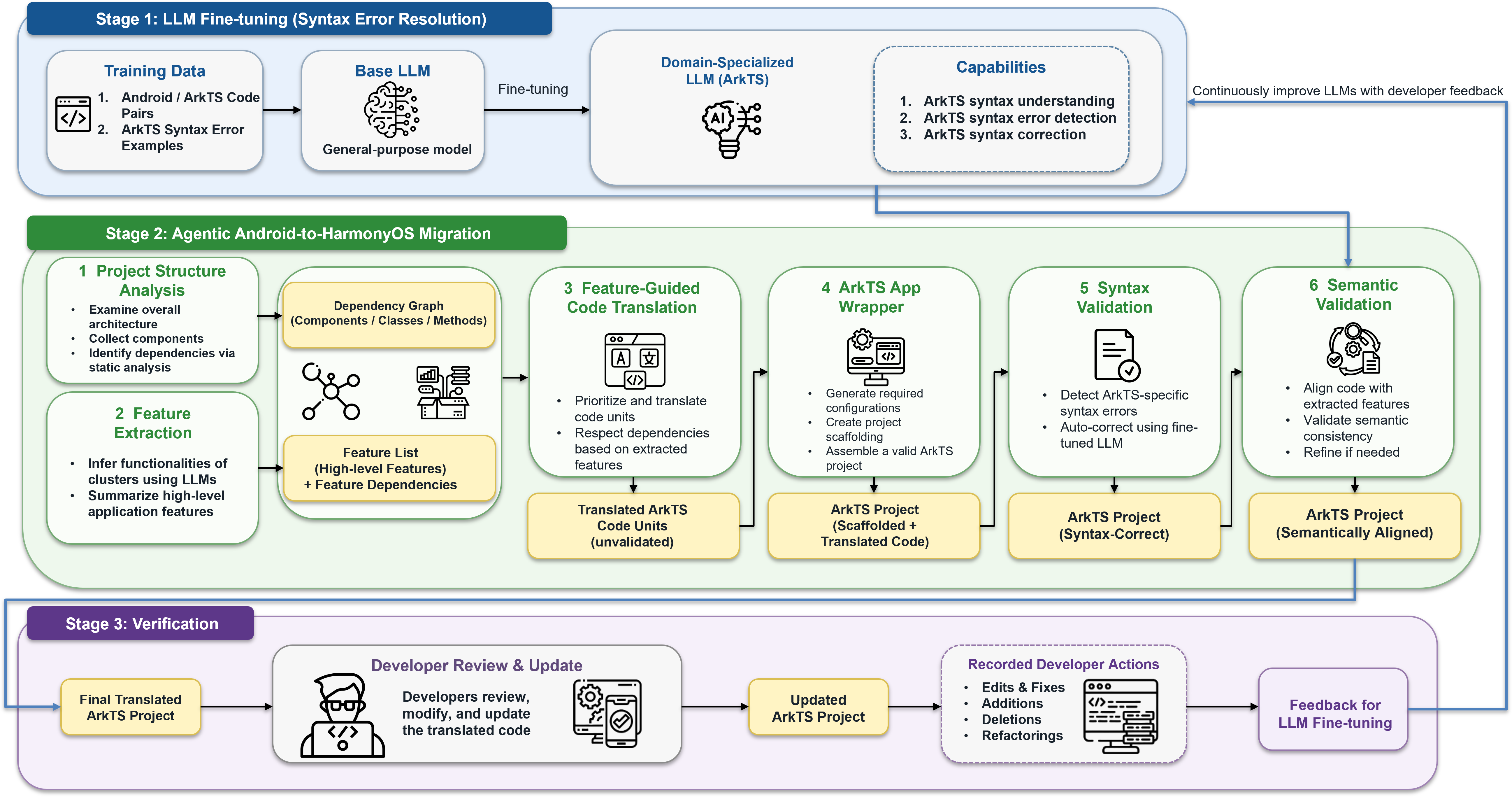}
  \caption{Overview of the proposed {\tool} framework}
  \label{fig:agent}
\end{figure}

Figure~\ref{fig:agent} illustrates the architecture of {\tool}, which consists of two main stages: LLM fine-tuning and Android-to-HarmonyOS migration.

\textbf{Stage1: LLM fine-tuning}.
This stage focuses on resolving syntax errors in ArkTS code generated by LLMs through fine-tuning a domain-specialized language model. This is motivated by our observation that general-purpose LLMs are ineffective at producing syntactically correct ArkTS code without targeted adaptation. 

\textbf{Stage2: Agentic Android-to-HarmonyOS
Migration}. This stage is further decomposed into six steps.  
Step \circled{1}: \textit{Project Structure Analysis} examines the overall architecture of an Android project, collects its components, and identifies dependencies at the component, class, and method levels through static analysis. 
Step \circled{2}: \textit{Feature Extraction} summarizes high-level application features by inferring the functionalities of these clusters using LLMs. 
Step \circled{3}: \textit{Feature-Guided Code Translation} prioritizes and translates code units while respecting 
dependencies based on the extracted features and their dependencies. 
Step \circled{4}: \textit{ArkTS App Wrapper} automatically generates a valid ArkTS project by creating the required configurations and scaffolding. After translation,
Step \circled{5}: \textit{Syntax Validation} leverages fine-tuned LLMs to detect and correct ArkTS-specific syntax errors, and Step \circled{6}: \textit{Semantic Validation} further refines the translated code by aligning it with extracted features and validating semantic consistency.

\textbf{Stage3: Manual Verification.} 
Finally, the translation results 
produced by {\tool} are presented to developers, who can review, modify, and update the translated code. All developer actions during this process are recorded and can be leveraged to further fine-tune the LLMs, improving their ability to generate correct ArkTS code. 

\subsection{Stage1: Fine-tuning LLMs}
\label{3.1}


We fine-tune relatively lightweight LLMs (e.g., 7B-16B parameters) to enable efficient local deployment for improved speed and practicality, following prior industry practices~\cite{murali2024ai}. Our goal is to develop specialized models to detect and correct syntax errors in LLM-generated ArkTS code. Due to the limited availability of high-quality ArkTS training data, we construct a carefully curated dataset of 1.7K specification–code pairs, extracted from the official ArkTS documentation and examples, to support fine-tuning. 

We adopt an instruction fine-tuning paradigm to train the specialized ArkTS-aware models. Each training instance is formulated as an instruction–response pair
consisting of (i) a specification describing the intended functionality and (ii) the corresponding ArkTS code example, both of which are directly extracted from official ArkTS documents (e.g., ArkTS Cookbook, ArkTS Standard Library, ArkTS
Specification, and ArkTS Tutorial).  All of these documents are publicly available.\footnote{\url{https://gitee.com/openharmony/arkcompiler_runtime_core/releases/tag/ArkTS-spec-1.2.0-alpha-20250307}}  
During fine-tuning, the model learns to follow task-oriented instructions and internalize ArkTS-specific syntax rules and language constructs, rather than relying on explicit syntactic error signals. 
The models are fine-tuned using supervised learning, optimizing the likelihood of the reference ArkTS code conditioned on the specification. 
This instruction fine-tuning process enables the model to acquire domain-specific knowledge and serve as a lightweight validator and corrector when applied in tandem with other LLMs in the translation pipeline.

Note that, in this paper, we currently rely solely on specification-code pairs extracted from the official ArkTS website for fine-tuning. In the future, user modifications and corrections made during the revision of translated code could also be collected and incorporated into the fine-tuning dataset, further improving the fine-tuned LLMs' ability to detect and correct ArkTS syntax errors.

\subsection{Stage2: Agentic Android-to-HarmonyOS Migration}
\label{sec:3.2}

\subsubsection{Project Structure Analysis} 

The first step in the Android-to-HarmonyOS migration is to analyze the Android project to be translated, creating a comprehensive code structure map that captures its architectural organization. It systematically decomposes the project into smaller, manageable units, i.e., components, classes, and methods, with static analysis. 
This step also collects detailed dependency information 
at the component level, class level, and method level, with static dependency analysis, identifying how different parts of the system interact and depend on one another. 
These analyses provide the foundation for subsequent translation tasks, enabling \tool\ to intelligently prioritize translation tasks. 

\subsubsection{Feature Extraction}

This step uses a general-purpose LLM (e.g., Claude-Sonnet-4.5) 
to perform a bottom-up summarization of the source code to identify and describe the system’s functional features.
It begins at the method level, where the intent of each function is inferred, capturing what a single method is designed to accomplish.
These method-level summaries are then aggregated to the class level, providing a holistic understanding of what functionality each class delivers. Finally, the process extends to the component level, where multiple classes are integrated to represent higher-level subsystem responsibilities. For each level-up summarization step, we concatenate the summaries from the immediate lower level into a single prompt and ask the LLM to synthesize a higher-level summary. 
This hierarchical semantic summarization enables the system to map low-level implementation details to high-level functionalities, 
which serve as essential guidance for feature-aligned translation and validation in subsequent stages.

Beyond code analysis, the module examines available project documentation (e.g., READMEs, release notes, and manual pages) to extract feature-related information from natural language descriptions, thereby improving the accuracy of feature understanding inferred from code.  

\subsubsection{Feature-Guided Code Translation}

This step employs a feature-guided, dependency-prioritized approach to code translation. Features with low dependency are translated first, as translating dependent features too early can cause the LLM to hallucinate, leading to missing interfaces or misbehaviours.  The prioritization algorithm is presented in Algorithm~\ref{alg:minimal-dependency-traversal}. 
Specifically, 
it begins by prioritizing features that involve the smallest number of components, as these are simpler and more self-contained. Within each feature, components with fewer classes are translated first to minimize interdependence. 
When a component contains multiple classes, the algorithm selects classes with the fewest dependencies, measured as the total number of other classes they depend on and the number of classes that depend on them. Here, a ``dependency'' is defined as an edge in the call graph. This ensures that classes with minimal interactions are processed first, reducing the risk of cascading errors during translation. 
Finally, within each class, method-level translations are ordered by their dependency levels.

Once the translation order is determined, {\tool} uses LLMs to translate the least dependent components, classes, or methods first to reduce the risk of cascading issues during translation. This hierarchical approach, from features to components, to classes, to methods, systematically reduces complexity and dependency propagation, enabling an incremental translation process.

\begin{algorithm}[t!]
\caption{Translation prioritization algorithm}
\label{alg:minimal-dependency-traversal}
\begin{algorithmic}[1]
\Require Features, Dependencies
\Ensure tor //TraversalOrder

\State $F \gets \Call{Sort}{Features,\ numberOfComponents}$ 
\Comment{\footnotesize Sort features by number of components (ascending)}
\State $tor \gets [\ ]$

\ForAll{$feature \in F$}
    \State $C \gets \Call{GetComponents}{feature}$
    \State $C' \gets \Call{Sort}{C,\ numberOfClasses}$
    \Comment{\footnotesize Sort components by number of classes (ascending)}
    
    \ForAll{$component \in C'$}
        \State $K \gets \Call{GetClasses}{component}$
        \State $K' \gets \Call{Sort}{K,\ classDependencyScore}$
        \Comment{\footnotesize Sort classes by dependency score (ascending)}
        
        \ForAll{$class \in K'$}
            \State $M \gets \Call{GetMethods}{class}$
            \State $M' \gets \Call{Sort}{M,\ methodDependencyScore}$
            \Comment{\footnotesize Sort methods by dependency score (ascending)}
            
 \ForAll{$method \in M'$}
\State $\Call{Append}{tor,( feature,component,class,method)}$
            \EndFor
        \EndFor
    \EndFor
\EndFor

\State \Return $tor$
\end{algorithmic}
\end{algorithm}

\subsubsection{ArkTS App Wrapper}

This step leverages LLMs to integrate the translated code from the previous step into a complete ArkTS project scaffold. Specifically, it uses a predefined structural template representing a typical ArkTS application layout as a one-shot prompt for the LLM. The template provides organized placeholders for key components, including the UI layout, component lifecycle hooks, permission handlers, and event bindings, ensuring a consistent and modular framework. During the LLM-based code update phase, the model takes the wrapper template, translated code, and manifest files as input, analyzes each logical unit, such as UI definitions, event handlers, and business logic, and maps them to their corresponding placeholders. This process lets the LLM automatically fill in and update the necessary sections, producing a cohesive, runnable ArkTS application that follows its architecture and coding conventions. 

\subsubsection{Syntax Validation}

In this step, the generated ArkTS code is systematically parsed and compiled using the official ArkTS compiler,\footnote{https://developer.huawei.com/consumer/en/arkcompiler/} which serves as an authoritative oracle for syntactic and semantic compliance. The compiler detects a wide range of syntax-related issues, such as incorrect or missing type annotations, absent semicolons, undeclared or improperly scoped variables, and incorrect usage of ArkTS-specific language constructs. 
When syntax errors are identified, the compiler-generated diagnostics are collected and paired with the corresponding code snippets that triggered the errors. This information is then fed back into the fine-tuned LLM described in Section~\ref{3.1} 
within a feedback-driven refinement loop. By explicitly grounding the model's revision process in concrete compiler feedback, the LLM is guided to generate targeted fixes. This iterative validation-and-correction mechanism continues until the code successfully passes syntax checking or reaches a predefined iteration limit. In this study, we conducted up to three rounds of iterative refinement for each compiler-detected issue. If the errors could not be resolved, {\tool} leaves them for human intervention. The maximum number of fix attempts was empirically determined to balance translation quality and generation overhead.

\subsubsection{Semantic Validation and Refinement}

This step focuses on ensuring that the generated ArkTS implementation faithfully preserves the original functionality of the source Android application. Rather than relying solely on syntactic validation, this stage explicitly targets semantic and behavioral equivalence between the original Android application and its ArkTS counterpart.

To achieve this, we employ an LLM-based alignment checker, referred to as the LLM-as-a-Judge, to evaluate the consistency between the original Android application and its ArkTS counterpart. The LLM-as-a-Judge is provided with artifacts from both implementations, including natural language feature specifications extracted and relevant code structure. Based on these inputs, the model performs a comparative analysis to determine whether the translated ArkTS code faithfully realizes the original feature requirements. 

During evaluation, the LLM-as-a-Judge identifies potential discrepancies such as deviations in logic, missing or incorrectly mapped event bindings, altered permission behaviors, or incomplete UI state transitions. It produces structured feedback that highlights both the nature and location of detected inconsistencies, and may optionally assign a confidence or alignment score reflecting the degree of equivalence. This feedback is then used to guide subsequent translation refinements in an iterative loop. In this paper, if the alignment score falls below 80\%, the generated code is refined using the LLM-as-a-Judge feedback 
until the threshold is met or a predefined iteration limit is reached, i.e., up to three rounds. 

\subsection{Stage3: Manual Verification}
\label{sec:3.3}



Finally, the translation results produced by {\tool} are presented to developers for human-in-the-loop validation. Developers can inspect, execute, and test the generated ArkTS code, and perform necessary modifications, refinements, or completions to ensure correctness, functionality, and adherence to best practices. This stage serves as a quality assurance step to catch subtle issues that may not be resolved automatically. Importantly, all developer interactions, such as edits, corrections, feedback, and acceptance decisions, are systematically logged. These interaction traces are then curated into high-quality supervision signals and can be leveraged to iteratively fine-tune the underlying LLMs in the future. 
\section{Case Study Evaluation}
\label{sec:4}

\begin{table*}[t!]
\centering
\caption{Overview of the Android projects used in our study.} 
\label{projects}
\begin{tabular}{llrrrrr}
\toprule
\textbf{Android Project} & \textbf{Description} & \textbf{\#Java} & \textbf{\#Classes} & \textbf{\#Packages} & \textbf{\#Methods} & \textbf{\#LOCs} \\
\midrule
Camera-picker              & Camera management   & 5   & 5   & 2   & 6     & 405 \\
Network-performance-timing & Network performance & 6   & 6   & 1   & 34    & 506 \\
Video-play                 & Video playback      & 13  & 11  & 4   & 123   & 1,278 \\
To\_Do\_List               & To-do list          & 13  & 13  & 1   & 113   & 2,005 \\
Materialistic              & News reader         & 131 & 131 & 8   & 1,686 & 28,303 \\
LeafPic                    & Photo gallery       & 139 & 137 & 25  & 1,705 & 32,342 \\
AndOTP                     & OTP manager         & 71  & 70  & 12  & 930   & 41,761 \\
Zxing                      & Barcode scanner     & 517 & 509 & 48  & 5,764 & 48,525 \\
Omni-Notes                 & Note manager        & 204 & 200 & 32  & 1,695 & 49,728 \\
Osmand                     & Offline navigation  & 172 & 353 & 17  & 4,297 & 60,227 \\
OpenTracks                 & Activity tracking   & 276 & 276 & 31  & 2,979 & 64,412 \\
Etar-Calendar              & Calendar manager    & 147 & 141 & 17  & 5,531 & 97,435 \\
AntennaPod                 & Podcast manager     & 609 & 593 & 106 & 6,494 & 114,012 \\
Open-keychain              & Key management      & 525 & 521 & 61  & 6,623 & 130,947 \\
\bottomrule
\end{tabular}
\end{table*}

\subsection{Research Questions}
\label{sec:4.1}

To systematically evaluate the capabilities and effectiveness of our approach, we focus on two key aspects of Android-to-HarmonyOS migration.

\noindent \textbf{RQ1: How does fine-tuning on ArkTS specification-code pairs improve translation accuracy?}

\vspace{4pt}
\noindent \textbf{RQ2: How effectively does {\tool} migrate Android projects to HarmonyOS?}

\vspace{4pt}
\noindent In \textbf{RQ1}, we focus on the impact of fine-tuned LLMs with a relatively small number of parameters (e.g., 7B–16B) on ArkTS code generation, examining improvements in translation accuracy and reductions in syntax errors. 
In \textbf{RQ2}, we evaluate the overall effectiveness of {\tool} in migrating Android projects to HarmonyOS, compared to two baselines. 




\subsection{Experiment Data}
\label{sec:4.2}

\subsubsection{Data For Fine-Tuning}
\label{sec:4.21}

All fine-tuning data used in this study were sourced exclusively from official HarmonyOS ArkTS documentation. 
We aggregated materials from four canonical sources: \textit{ArkTS Cookbook}, \textit{ArkTS Standard Library}, \textit{ArkTS Specification}, and \textit{ArkTS Tutorial}. All of these documents are publicly available online (as discussed in Section~\ref{3.1}). 
These sources constitute the most authoritative references provided by the platform vendor and comprehensively cover standard library definitions, core language semantics, idiomatic usage patterns, and pedagogical examples.

To convert these documents into a supervised fine-tuning dataset, we first develop an automated script to extract example code embedded in the official materials and align each code snippet with its corresponding natural-language explanation provided in the documents. For each documented API, language construct, or usage scenario, the associated textual description typically specifies the function purpose, expected behavior, parameter semantics, and usage constraints, and is treated as the model input, while the authoritative ArkTS implementation provided in the documentation is used as the output. This process yields structured input–output pairs that map natural-language described functionalities 
directly to their canonical ArkTS reference implementations, enabling the model to learn how formal semantic descriptions translate into executable ArkTS code. To ensure the quality and correctness of the automatically extracted data, three of the authors collaboratively reviewed every data instance to verify semantic alignment between descriptions and code and to remove noisy, incomplete, or ambiguous examples. The final dataset consists of 1,752 high-quality pairs.

\subsubsection{Testing Data For Fine-Tuning}
\label{sec:4.22}

To evaluate the performance of our fine-tuned LLMs, we manually curate 
a dataset consisting of 200 standalone functions sourced from the AVATAR dataset~\cite{ahmad2023avatar}, which has been widely adopted for benchmarking function-level program translation between Java and Python~\cite{ibrahimzada2025alphatrans,yuan2024transagent,pan2024lost,rabbi2025babelcoder}. AVATAR is a large-scale collection of 9,515 programming problems along with their solutions in both Java and Python. The dataset is compiled from competitive programming platforms, online coding platforms, and open-source repositories, and includes unit tests for 250 examples, enabling systematic evaluation of functional correctness. For our experiments, we select these Java programs that include unit tests and convert them into ArkTS translation tasks, preserving and adapting the unit tests to verify correctness.

\subsubsection{Android Projects for Translation}
\label{sec:4.23}

To evaluate the effectiveness of our approach for translating Android applications to HarmonyOS, we curate 
a diverse set of ten Android applications across multiple domains and usage scenarios. These projects were selected to reflect a broad range of real-world mobile applications while remaining tractable for detailed manual inspection and analysis. Specifically, we prioritized repositories covering different application domains, such as to-do list applications, media players, and utility apps, to ensure that our evaluation captures a diverse set of functionalities and interaction patterns common in Android apps. 
Importantly, we excluded projects that heavily depend on platform-specific third-party libraries, such as Google Play Services, to minimize external confounding factors and focus the evaluation on core application logic and platform-dependent code. This selection strategy enables a more accurate assessment of the capabilities and limitations of {\tool} in handling essential mobile development tasks. In addition, we excluded projects implemented in Kotlin, as our current approach targets Java-based Android applications. 
Table~\ref{projects} summarizes the key characteristics of the selected Android applications. 

\subsection{Base LLMs}
\label{sec:4.3}

\subsubsection{LLMs for Fine-tuning}
\label{subsubsec:llms-for-fine-tuning}
To fine-tune models for automated syntax error correction, we select three instruction-tuned large language models specialized for code understanding and generation, i.e., \textbf{Qwen2.5-Coder-7B-Instruct\footnote{https://huggingface.co/Qwen/Qwen2.5-Coder-7B-Instruct}}, \textbf{DeepSeek-Coder-V2-Lite-16B-Instruct\footnote{https://huggingface.co/deepseek-ai/DeepSeek-Coder-V2-Lite-Instruct}}, and \textbf{CodeLLaMA-13B-Instruct\footnote{https://huggingface.co/codellama/CodeLlama-13b-Instruct-hf}}. 
These models were chosen to represent diverse architectural designs and parameter scales, allowing us to examine the effectiveness of syntax repair across different model capacities. All three models are fine-tuned using the same dataset and training protocol to ensure fair comparisons across models.

\subsubsection{LLMs for Running {\tool}}

We experiment with four LLMs, i.e., \textbf{GPT-4.1}, \textbf{Claude-Sonnet-4.5}, \textbf{Qwen3-VL-235B-A22B-Instruct}, and \textbf{ Gemini-2.5-Flash}, provided by OpenRouter\footnote{https://openrouter.ai/}, to drive our agentic Android-to-ArkTS migration pipeline.

Note that we do not compare these models using the full end-to-end translation workflow. This decision is primarily motivated by practical considerations: the complete pipeline requires a large number of LLM API calls and substantial human effort to validate and manually refine the final translated applications, making it difficult to conduct fair and reproducible comparisons across models. 
Instead, we adopt a simplified end-to-end evaluation setting, focusing on direct Java-to-ArkTS translation on the curated test dataset (See Section~\ref{sec:4.23}). This controlled setup allows us to compare models under identical conditions while minimizing external variability introduced by iterative refinement and human intervention. The results of this comparison are then used to select the most suitable model to drive our agentic translation approach. For all the models, we set the temperature set to zero to ensure a more deterministic and reliable response~\cite{shin2024retrieval}.


\subsection{Evaluation Metrics}

To evaluate the performance of fine-tuned LLMs, we follow prior work and adopt the $Pass@1$ metric~\cite{chen2021evaluating}. Pass@1 quantifies function-level translation accuracy as the proportion of generated translations that pass all unit tests on the first attempt, thereby capturing the model’s precision and reliability in direct Java-to-ArkTS code translation.




To comprehensively assess the agentic translation framework, we employ a human-in-the-loop evaluation protocol. In this process, developers manually inspect the translated applications and assess the quality of their user interfaces and behavioral correctness across the following dimensions:

\begin{itemize}

\item \textbf{$Functionality\_Completeness$}: This metric measures the extent to which application features are migrated from the Android project to the ArkTS project, regardless of whether they are fully or partially functional. It evaluates feature-level coverage from a user perspective, focusing on whether key features are present in the migrated application. A feature receives a high score if its primary entry points and essential interaction flows are implemented; it receives a low score if it is missing, only stubbed, or significantly reduced in scope. 

\item  \textbf{$Functionality\_Correctness$}: This metric evaluates whether the migrated functionality is correctly implemented. It assesses whether a feature, once triggered, produces the expected outcomes, state transitions, and outputs, consistent with the original Android behavior when an equivalent or appropriately adapted ArkTS implementation is feasible. 


\item  \textbf{$UI\_Operability$}: This metric evaluates how easily users can discover, navigate, and operate the migrated application. Reviewers assess factors such as UI responsiveness, clarity of navigation, consistency of controls, back-navigation behavior, and whether the interaction flow is intuitive without inspecting the underlying code. A high score indicates a smooth, coherent, and user-friendly experience with well-integrated features, while a low score reflects issues such as confusing navigation, unresponsive elements, dead-end pages, or broken interactions. 

\item \textbf{$UI\_Alignment$}: This metric measures how closely the migrated ArkTS application's UI aligns with the original Android application's user-visible interface design. It compares layout structure, information hierarchy, visual organization, interaction style, and overall presentation. A high score indicates that the migrated UI preserves the original screen organization and provides a similar user experience; a low score indicates significant visual divergence, missing interface regions, inconsistent layout behavior, or a substantially different interaction experience.

\item \textbf{$Time\_Cost$}: We also record the time required by each approach to translate an Android project into ArkTS, measured from the start of translation to the generation of a deployable project. 


\end{itemize}

All metrics except \textbf{$Time\_Cost$} are scored on a scale ranging from 0 to 100, where a higher score indicates better performance with respect to the evaluated dimension. 

We incorporate manual UI and behavior review because automated metrics alone cannot capture the true quality of migrated applications. Many failures are semantic rather than syntactic: apps may compile and appear complete, yet fail to support core user tasks due to broken navigation, missing functionality, or inconsistent behavior. Automated metrics (e.g., file counts or compile success) may overestimate the prevalence of such cases. Manual inspection allows us to assess end-to-end usability, including workflow coherence and meaningful interaction, complementing automated measures.

\subsection{Baseline Approach}

In this paper, we use a basic LLM-based prompting approach and the Claude Code CLI~\cite{claude} as the baselines to evaluate the performance of {\tool}. 
To ensure a fair comparison, we use the same underlying large language model, Claude-Sonnet-4.5, to drive {\tool}, the prompt-based approach, and Claude Code CLI. 

For the prompting baseline, we design structured prompts that explicitly describe the Android-to-ArkTS migration task. The prompts guide the model to understand the source Android application, including its features, user interfaces, components, classes, and functionalities, and then generate the corresponding ArkTS implementation. They also instruct the model to preserve the original application functionality and UI behavior during migration. To ensure comparability with {\tool}, we further adopt a similar verification and refinement loop, in which the generated ArkTS code is compiled, compiler errors are collected, and the model is prompted to iteratively revise the implementation until the code passes compilation. 

For the Claude Code CLI baseline, we implement a task-specific skill that performs agentic Android-to-ArkTS migration using its built-in tools. The skill analyzes the Android project structure and incrementally translates it into ArkTS components and HarmonyOS APIs. It also adopts a verification–refinement loop, where generated ArkTS code is compiled, errors are collected, and the model iteratively revises the code until successful compilation.


\section{Result Analysis}
\label{sec:result}

\subsection{RQ1: Effectiveness of Fine-Tuned LLMs}
\label{sec:5.11}

\usepgfplotslibrary{groupplots}
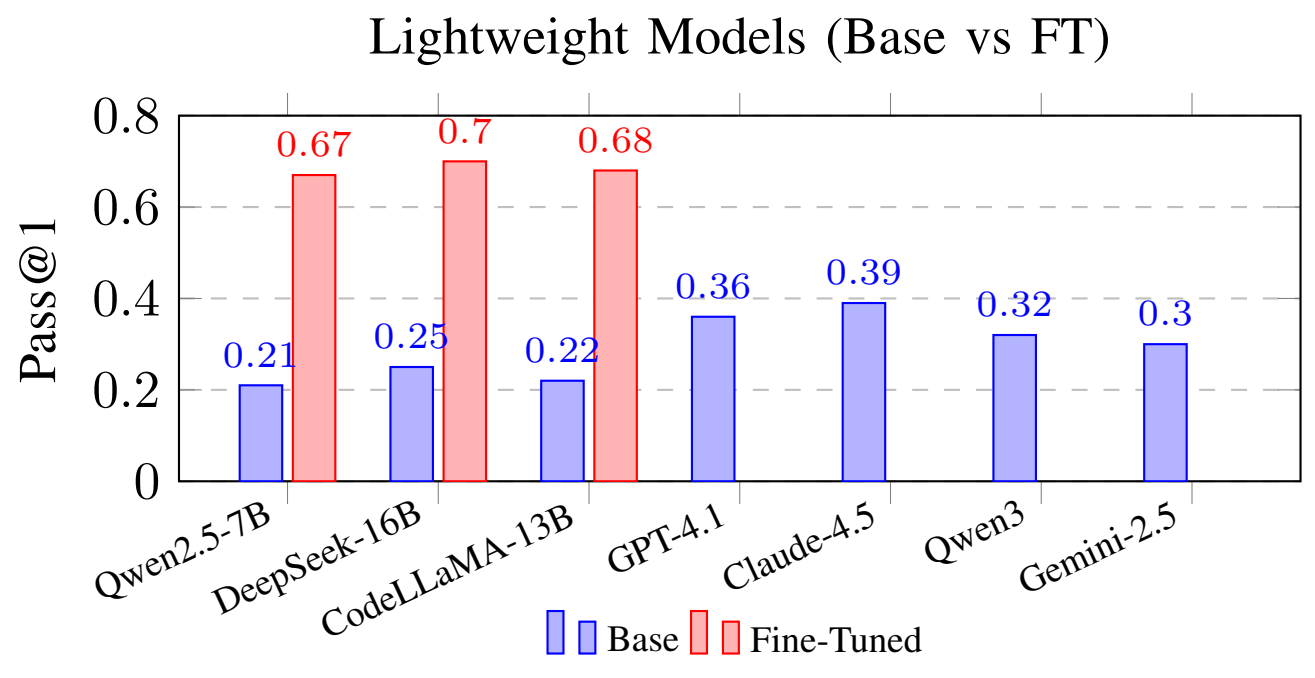
\begin{figure}[t!]
\centering
\begin{tikzpicture}
\begin{axis}[
  ybar,
  width=9cm,
  height=4cm,
  ymin=0,
  ymax=0.8,
  bar width=8pt,
  enlarge x limits=0.12,
  ylabel={Pass@1},
  symbolic x coords={
    Qwen2.5-7B,
    DeepSeek-16B,
    CodeLLaMA-13B,
    GPT-4.1,
    Claude-4.5,
    Qwen3,
    Gemini-2.5
  },
  xtick=data,
  x tick label style={
    rotate=25,
    anchor=east,
    font=\scriptsize
  },
  ymajorgrids=true,
  grid style=dashed,
  nodes near coords,
  nodes near coords style={font=\scriptsize},
  nodes near coords align={vertical},
  title={Lightweight Models (Base vs FT)},
  legend style={
    font=\scriptsize,
    at={(0.5,-0.32)},
    anchor=north,
    legend columns=2,
    draw=none,
    fill=none
  },
]

\addplot coordinates {
  (Qwen2.5-7B, 0.21)
  (DeepSeek-16B, 0.25)
  (CodeLLaMA-13B, 0.22)
  (GPT-4.1, 0.36)
  (Claude-4.5, 0.39)
  (Qwen3, 0.32)
  (Gemini-2.5, 0.30)
};

\addplot coordinates {
  (Qwen2.5-7B, 0.67)
  (DeepSeek-16B, 0.70)
  (CodeLLaMA-13B, 0.68)
};

\legend{Base, Fine-Tuned}

\end{axis}
\end{tikzpicture}
\caption{Performance of Base and Fine-Tuned Models on Java-to-ArkTS Translation.}
\label{fig:pass1}
\end{figure}

\noindent \textbf{Approach:} We adopt a parameter-efficient LoRA-based strategy to fine-tune the three selected models for syntax error correction (Section~\ref{subsubsec:llms-for-fine-tuning}). LoRA (Low-Rank Adaptation) updates only a subset of model parameters while keeping the majority of the pretrained weights frozen. This approach significantly reduces GPU memory usage and training time while enabling the model to learn task-specific behaviors. Following existing work~\cite{weyssow2025exploring}, in our experiments, we adopt low-rank adaptation matrices with a rank of 16 ($r=16$) and a scaling factor of 32 ($lora\_alpha=32$). To prevent overfitting and improve generalization, a dropout rate of $0.05$ ($lora\_dropout=0.05$) is applied to the adaptation weights. All fine-tuning experiments are conducted on a single NVIDIA A100 SXM4 GPU. 
We fine-tune the models using the dataset described in Section~\ref{sec:4.21}, and subsequently evaluate their performance on the Java-to-ArkTS translation dataset introduced in Section~\ref{sec:4.22}. To assess translation effectiveness, we adopt the $pass@1$ metric.

\noindent \textbf{Results:} The detailed $pass@1$ of the base models and the fine-tuned models are shown in Fig.~\ref{fig:pass1}. 
As we can see, fine-tuning on this specific translation task significantly boosts performance: Qwen2.5-7B-FT, CodeLLaMA-13B-FT, and DeepSeek-16B-FT achieve $pass@1$ scores of 0.67, 0.68, and 0.70, respectively, representing roughly a threefold improvement over their base counterparts. In addition, the fine-tuned models substantially outperform the LLMs with a larger parameter size, i.e., GPT-4.1 (0.36), Claude-Sonnet-4.5 (0.33), Qwen-3 (0.32), and Gemini-2.5 (0.30), demonstrating the critical importance of fine-tuning for low-resource programming language translation tasks. 

\mybox{\textbf{Answer to RQ1:} Experimental results indicate that, for Java-to-ArkTS translation, careful task-specific fine-tuning can significantly improve LLM performance in generating correct and syntactically valid ArkTS code.}

\subsection{RQ2: Effectiveness of {\tool}}
\label{sec:5.12}

\begin{table*}[t]
\centering
\setlength{\tabcolsep}{2.5pt}
\caption{Comparison of {\tool} with the two baselines. Comp, Corr, Appr, Oper, and Align denote functional completeness, functional correctness, functional appropriateness, operability, and UI alignment, respectively. `T' is the time cost.}
\label{tab:manual-ui-behavior-func-ui}
\resizebox{\textwidth}{!}{%
\begin{tabular}{lrrrrrrrrrrrrrrrrrr}
\toprule
Project & \multicolumn{6}{c}{Prompt-based} & \multicolumn{6}{c}{Claude Code} & \multicolumn{6}{c}{A2HTrans} \\
\cmidrule(lr){2-7}\cmidrule(lr){8-13}\cmidrule(lr){14-19}
& Comp & Corr & Appr & Oper & Align & T(min) & Comp & Corr & Appr & Oper & Align & T(min) & Comp & Corr & Appr & Oper & Align & T(min) \\
\midrule
camera-picker-master & \textbf{100.0} & \textbf{100.0} & \textbf{100.0} & \textbf{100.0} & \textbf{100.0} & 3.95 & \textbf{100.0} & \textbf{100.0} & \textbf{100.0} & \textbf{100.0} & \textbf{100.0} & 7.27 & \textbf{100.0} & \textbf{100.0} & \textbf{100.0} & \textbf{100.0} & \textbf{100.0} & \textbf{3.00} \\
NetworkPerformanceTiming & \textbf{100.0} & \textbf{100.0} & \textbf{100.0} & \textbf{100.0} & \textbf{100.0} & 6.23 & \textbf{100.0} & \textbf{100.0} & \textbf{100.0} & \textbf{100.0} & \textbf{100.0} & 7.91 & \textbf{100.0} & \textbf{100.0} & \textbf{100.0} & \textbf{100.0} & \textbf{100.0} & \textbf{3.21} \\
videoplayback & 82.5 & 77.5 & 82.5 & 92.5 & \textbf{92.5} & 16.31 & \textbf{87.5} & \textbf{82.5} & \textbf{85.0} & 90.0 & \textbf{92.5} & 12.03 & \textbf{87.5} & \textbf{82.5} & \textbf{85.0} & \textbf{95.0} & \textbf{92.5} & \textbf{4.22} \\
To\_Do\_List-master & \textbf{97.5} & \textbf{97.5} & \textbf{95.0} & 95.0 & 92.5 & 18.09 & 87.5 & 82.5 & 90.0 & 92.5 & 90.0 & 6.69 & 92.5 & 95.0 & \textbf{95.0} & \textbf{97.5} & \textbf{97.5} & \textbf{4.89} \\
materialistic & 87.5 & 90.0 & 90.0 & 92.5 & 90.0 & 148.62 & 47.5 & 45.0 & 57.5 & 50.0 & 62.5 & 16.42 & \textbf{92.5} & \textbf{95.0} & \textbf{92.5} & \textbf{95.0} & \textbf{92.5} & \textbf{10.09} \\
LeafPic & 45.0 & 62.5 & 62.5 & 60.0 & 62.5 & 150.88 & 75.0 & 72.5 & \textbf{82.5} & 75.0 & 72.5 & 16.71 & \textbf{82.5} & \textbf{75.0} & 80.0 & \textbf{77.5} & \textbf{87.5} & \textbf{10.01} \\
andOTP & 50.0 & 70.0 & 55.0 & 80.0 & 55.0 & 89.29 & 72.5 & 67.5 & \textbf{87.5} & 90.0 & \textbf{82.5} & 24.96 & \textbf{85.0} & \textbf{77.5} & \textbf{87.5} & \textbf{92.5} & \textbf{82.5} & \textbf{12.36} \\
zxing & 57.5 & \textbf{67.5} & 55.0 & 67.5 & 67.5 & 169.13 & 65.0 & \textbf{67.5} & \textbf{82.5} & \textbf{87.5} & \textbf{85.0} & 16.26 & \textbf{75.0} & 65.0 & \textbf{82.5} & 82.5 & \textbf{85.0} & \textbf{11.86} \\
Omni-Notes & 70.0 & \textbf{70.0} & 77.5 & 75.0 & 77.5 & 171.24 & \textbf{80.0} & 65.0 & \textbf{82.5} & \textbf{80.0} & 80.0 & 14.57 & \textbf{80.0} & 67.5 & 80.0 & \textbf{80.0} & \textbf{87.5} & \textbf{11.46} \\
Osmand & 50.0 & 55.0 & 50.0 & 60.0 & 62.5 & 264.75 & \textbf{82.5} & \textbf{77.5} & \textbf{85.0} & \textbf{80.0} & \textbf{85.0} & \textbf{20.13} & 72.5 & 70.0 & 77.5 & 77.5 & 80.0 & 20.27 \\
OpenTracks & 52.5 & 60.0 & 65.0 & 75.0 & 75.0 & 182.90 & 70.0 & 72.5 & 70.0 & \textbf{87.5} & 80.0 & 17.03 & \textbf{80.0} & \textbf{77.5} & \textbf{75.0} & \textbf{87.5} & \textbf{92.5} & \textbf{11.79} \\
Etar-Calendar & 70.0 & \textbf{72.5} & \textbf{85.0} & \textbf{85.0} & \textbf{90.0} & 128.74 & 70.0 & 47.5 & 72.5 & 72.5 & 80.0 & \textbf{17.30} & \textbf{75.0} & 62.5 & \textbf{85.0} & 80.0 & \textbf{90.0} & 19.87 \\
AntennaPod & 40.0 & 70.0 & 50.0 & 57.5 & 47.5 & 192.12 & 65.0 & \textbf{72.5} & \textbf{82.5} & 80.0 & \textbf{75.0} & \textbf{16.49} & \textbf{72.5} & \textbf{72.5} & \textbf{82.5} & \textbf{85.0} & 70.0 & 18.43 \\
open-keychain & 52.5 & 55.0 & 57.5 & 62.5 & 70.0 & 214.65 & \textbf{82.5} & \textbf{85.0} & \textbf{90.0} & \textbf{87.5} & \textbf{95.0} & 23.69 & 80.0 & 80.0 & 80.0 & 85.0 & 87.5 & \textbf{13.41} \\
\midrule
Avg. & 68.2 & 74.8 & 73.2 & 78.8 & 77.3 & 125.49 & 77.5 & 74.1 & 83.4 & 83.8 & 84.3 & 15.53 & \textbf{83.9} & \textbf{80.0} & \textbf{85.9} & \textbf{88.2} & \textbf{88.9} & \textbf{11.06} \\
\bottomrule
\end{tabular}%
}
\end{table*}

\noindent \textbf{Approach:}
Based on our preliminary experimental results in RQ1 (see Section~\ref{sec:5.11}), we select the fine-tuned DeepSeek-16B-Instruct (DeepSeek-16B-FT) model for the syntax validation stage, as it achieves the highest $pass@1$ performance and demonstrates strong robustness in generating syntactically correct ArkTS code. In this stage, DeepSeek-16B-FT primarily detects and corrects syntax errors in the translated ArkTS programs. For the agentic migration process from Android applications to their ArkTS counterparts, 
we used Claude-Sonnet-4.5 
as the driving model of {\tool} for the Android-to-ArkTS Migration. Claude-Sonnet-4.5 is chosen for its strong general reasoning and planning capabilities, which are essential for coordinating multi-step translation actions, such as analyzing Android project structures, mapping application components, and orchestrating iterative interactions among translation, validation, and refinement modules. It also outperforms  GPT-4.1, Qwen-3, and Gemini-2.5 in the Java-to-ArkTs translation task, as shown in Figure~\ref{fig:pass1}. 
For a fair comparison, we also use Claude Sonnet 4.5 for both baselines, i.e., the prompt-based approach and the Claude Code CLI. 

We experiment with the 14 projects using the metrics \textbf{$Functionality$} \textbf{$\_Completeness$}, \textbf{$Functionality\_Correctness$}, \textbf{$UI\_Operability$}, and \textbf{$UI\_Alignment$} to measure the effectiveness of {\tool}. To obtain reliable evaluations, two authors manually reviewed the translations generated by {\tool} and two baseline approaches. Each participant independently examines the translated results and assigns scores for all four metrics based on predefined criteria. The final score for each metric is computed by averaging the ratings across the three participants, thereby reducing individual bias and improving evaluation robustness.

\noindent \textbf{Results}: 
Table~\ref{tab:manual-ui-behavior-func-ui} presents a comparison of {\tool} with the prompt-based approach and Claude Code across functionality, UI quality, and time cost. Overall, {\tool} consistently achieves the best performance across all metrics. 
First, regarding functionality, {\tool} achieves the highest average scores across all three functionality-oriented metrics: functional completeness, functional correctness, and functional appropriateness, with scores of 83.9\%, 80.0\%, and 85.9\%, respectively. In comparison, the prompt-based approach achieves 68.2\%, 74.8\%, and 73.2\%, while Claude Code achieves 77.5\%, 74.1\%, and 83.4\%. These results indicate that {\tool} not only preserves a larger portion of the expected functionality, but also better maintains correct behavior and task-level appropriateness in the migrated HarmonyOS applications. 
Second, regarding UI quality, {\tool} achieves the highest average scores on both UI-related metrics, with 88.2\% in operability and 88.9\% in UI alignment. In comparison, the prompt-based approach achieves 78.8\% and 77.3\%, while Claude Code achieves 83.8\% and 84.3\%. These results show that {\tool} better preserves user interaction quality and interface consistency with the original Android applications.
Third, regarding time cost, {\tool} is the most efficient method, with an average runtime of 11.06 minutes. In comparison, the prompt-based approach requires 125.49 minutes, while Claude Code requires 15.53 minutes. This indicates that {\tool} improves migration quality while also reducing translation time.

\begin{figure*}[t]
  \centering
  \includegraphics[width=0.8\textwidth]{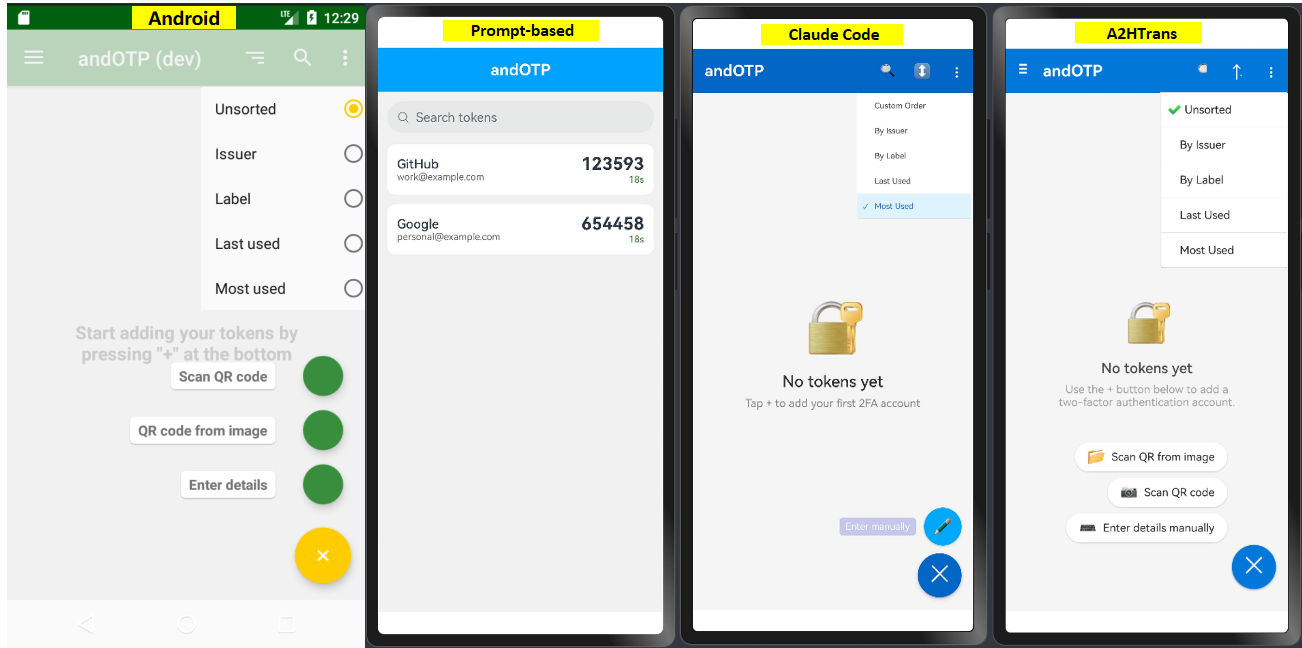}
  \caption{Case study on migrating the \textsc{andOTP} application. From left to right: the original Android app, a prompt-based baseline, Claude Code, and {\tool}. The prompt-based baseline misses key interactions, and Claude Code restores the UI only partially while failing to preserve essential workflows. {\tool} best preserves both the interface structure and user interactions, resulting in higher behavioral fidelity.}
  \label{fig:andOTP_example}
\end{figure*}

Figure~\ref{fig:andOTP_example} presents a case study illustrating UI-level differences between {\tool} and the two baselines. In the original Android app, the startup flow exposes several key actions on the empty-token screen, including a sorting menu and three add-token options: scanning a QR code, importing a QR image, and manual entry. 
The prompt-based approach reproduces the token list UI but fails to capture these essential interaction structures. Claude Code reconstructs the overall token-list interface and includes a sorting menu; however, it compresses the add-token workflow into a limited manual-entry option, failing to fully recover the original entry actions. In contrast, {\tool} preserves the empty-state structure while exposing all three token-entry options in a manner closely aligned with the original Android design. 
This case highlights the importance of manual, behavior-level evaluation: although all generated apps present a plausible empty token list, they differ substantially in whether the startup screen supports the full range of task-level interactions expected by the original application. 

We provide a comparison of the UIs of applications translated by {\tool} and the two baselines for each of the 14 projects in our replication package\footnote{\url{https://anonymous.4open.science/r/A2HTrans-9F33/README.md}}.

\mybox{\textbf{Answer to RQ2:} {\tool} outperforms both baselines across all metrics, achieving higher functionality completeness and correctness, superior UI operability and alignment, and the lowest time cost. It produces more accurate and consistent translations while being significantly more efficient than the prompt-based approach and Claude Code, demonstrating overall effectiveness and practicality.}

\section{Lesson Learned from Translation Issues}
\label{sec:lesson}

\begin{table}[t]
\centering
\caption{Overall quality-issue categories in agent-based Android-to-ArkTS migration.} 
\label{tab:claude-migration-weakness-overall}
\begin{tabular}{lrr}
\toprule
Quality-Issue Category & Count & Share \\
\midrule
\textbf{Scope contraction} & \textbf{194} & \textbf{25.9\%} \\
\multicolumn{1}{l}{\quad Complex Android UI pattern} & \multicolumn{1}{r}{\quad 69} & \multicolumn{1}{r}{\quad 35.6\%} \\
\multicolumn{1}{l}{\quad Other feature omission} & \multicolumn{1}{r}{\quad 37} & \multicolumn{1}{r}{\quad 19.1\%} \\
\multicolumn{1}{l}{\quad Media/hardware/external I/O} & \multicolumn{1}{r}{\quad 26} & \multicolumn{1}{r}{\quad 13.4\%} \\
\multicolumn{1}{l}{\quad Android ecosystem dependency} & \multicolumn{1}{r}{\quad 18} & \multicolumn{1}{r}{\quad 9.3\%} \\
\multicolumn{1}{l}{\quad Security/crypto/trust layer} & \multicolumn{1}{r}{\quad 14} & \multicolumn{1}{r}{\quad 7.2\%} \\
\multicolumn{1}{l}{\quad Data sync/persistence} & \multicolumn{1}{r}{\quad 11} & \multicolumn{1}{r}{\quad 5.7\%} \\
\textbf{Build scaffold constraint} & \textbf{166} & \textbf{22.2\%} \\
\textbf{Unknown or unclear} & \textbf{89} & \textbf{11.9\%} \\
\textbf{Platform API gap} & \textbf{76} & \textbf{10.2\%} \\
\textbf{Data persistence gap} & \textbf{71} & \textbf{9.5\%} \\
\textbf{Hardware/sensor/external I/O gap} & \textbf{60} & \textbf{8.0\%} \\
\textbf{Shallow UI shell} & \textbf{38} & \textbf{5.1\%} \\
\textbf{Third-party dependency gap} & \textbf{36} & \textbf{4.8\%} \\
\textbf{Routing/startup gap} & \textbf{18} & \textbf{2.4\%} \\
\midrule
\textbf{Total} & \textbf{748} & \textbf{100.0\%} \\
\bottomrule
\end{tabular}
\end{table}

To better understand why feature preservation remains challenging and what we can learn from these issues for future cross-platform translation techniques and tool design, we analyzed translation issues in the generated ArkTS projects using an open-coding procedure. Two authors independently inspected each {\tool} output across multiple evidence sources, including the generated ArkTS source code, runtime application behavior, and repair records (e.g., documented build failures). Disagreements were discussed and resolved through consensus.

The authors first assigned descriptive labels to observed issues and iteratively consolidated recurring labels into a fixed taxonomy. The final taxonomy captures failures at different layers of the translation process: \textit{scope contraction} refers to reduced feature coverage in the migrated app; \textit{build scaffold constraint} captures incomplete or incorrect project configuration, metadata, or dependency structure; \textit{platform API gap} includes unsupported Android-specific mechanisms (e.g., intents, services, widgets, lifecycle handling); \textit{data persistence gap} reflects missing storage components such as databases, repositories, caches, preferences, or synchronization logic; \textit{hardware/sensor gap} involves incomplete integrations with components like camera, GPS, media store, file pickers, or NFC; \textit{shallow UI shell} denotes UI elements that exist structurally but lack meaningful or correct behavior; \textit{third-party dependency gap} refers to missing or incompatible external libraries; and \textit{routing/startup gap} captures issues in entry points, page registration, or navigation flows. After establishing the taxonomy, each identified issue was assigned a single primary category to enable consistent analysis.


Table~\ref{tab:claude-migration-weakness-overall} reports the resulting quality-issue distribution by issue count. In total, 748 translation issues were observed and analyzed across the migrated projects. The most prevalent category is \textit{scope contraction} (194 issues, 25.9\%), indicating that many translated apps lose important functionality, particularly complex UI patterns and feature completeness. This is followed by \textit{build scaffold constraint} (166 issues, 22.2\%), highlighting difficulties in reconstructing correct project configurations and dependencies. 
\textit{platform API gap} (76 issues, 10.2\%) and \textit{data persistence gap} (71 issues, 9.5\%) further show challenges in handling Android-specific mechanisms and maintaining application state. Issues related to hardware/sensor/external I/O (60 issues, 8.0\%) also remain significant. 
At the UI level, \textit{shallow UI shell} (38 issues, 5.1\%) suggests that interfaces are often superficially generated without full functionality. \textit{Third-party dependency gap} (36 issues, 4.8\%) and \textit{routing/startup gap} (18 issues, 2.4\%) are less frequent but still impactful, affecting app completeness and usability.


\textbf{Implications}:
These quality issues suggest that Android-to-ArkTS migration cannot be treated as a direct code translation problem. Many failures arise because Android/Java and HarmonyOS/ArkTS differ fundamentally in application architecture, not just syntax. Android applications distribute behavior across UI components, lifecycle callbacks, platform services, build configurations, resources, persistence layers, hardware interfaces, and third-party libraries. In contrast, HarmonyOS relies on a different project model, ArkUI’s declarative component hierarchy, JSON5-based configuration, and distinct routing and lifecycle mechanisms. As a result, direct translation may preserve code fragments or surface-level UI, yet still lose feature scope, structural integrity, platform integration, or complex runtime behavior. This explains why migrated applications may appear plausible but fail to support complete task-level functionality. 

To improve migration quality, future systems should treat migration as an architectural adaptation problem rather than a file-level translation task. A practical workflow should first recover the source application’s feature intent and dependency structure, then map each feature to an appropriate HarmonyOS design before generating ArkTS code. Platform-knowledge retrieval can help identify equivalent HarmonyOS APIs or determine when no direct replacement exists, while dependency feasibility checks can prevent hallucinated libraries or unsupported Android components. 
Also, the lack of automated approaches for evaluating applications across different versions calls for more robust and standardized evaluation frameworks. In practice, it remains difficult to systematically compare multiple translated versions of an app in terms of functionality, behavior, and UI consistency without relying on manual inspection. This limits scalability and reproducibility. Future work should explore automated testing, interaction-based evaluation, and benchmark suites that can capture task-level behavior, state transitions, and user workflows across versions, enabling more consistent and scalable assessment of migration quality.

\section{Threats to Validity}
\label{sec:threats}

We evaluate 14 real-world Android applications spanning different domains, sizes, and architectural patterns, which improves diversity. However, this sample may still not fully represent the broader Android app ecosystem, particularly large-scale industrial systems, highly specialized domains, or apps with complex backend integrations. In addition, our results depend on the specific LLMs and tool configurations used in this study. Different base models, future model updates, or alternative prompting, translation, and repair strategies may yield different outcomes. Therefore, while our findings indicate strong performance, they may not generalize uniformly across all settings.

The fine-tuning dataset consists exclusively of publicly available ArkTS examples from official documentation. Although these sources are authoritative, they may not capture the full diversity and complexity of real-world ArkTS development, potentially limiting the representativeness of the learned semantics. Furthermore, our evaluation relies on manual assessments conducted by three developers with experience in both Android and ArkTS. While we mitigate individual bias through multiple reviewers and score averaging, some degree of subjectivity remains. 
\section{Conclusion}
\label{sec:conclusion}

In this paper, we presented {\tool}, an agentic framework for migrating Android applications to HarmonyOS/ArkTS through feature-guided analysis, project-level reconstruction, and ArkTS-specific repair. Evaluation on 14 real-world Android applications shows that {\tool} effectively supports the automatic migration of Android applications to HarmonyOS. Compared to Claude Code, {\tool} preserves 8.3\% more functional completeness and improves behavioral correctness by 8.0\%, while reducing translation time by 28.8\%. 
Our analysis of quality issues shows that the primary challenge lies not in isolated syntax translation, but in the architectural mismatch between Android and HarmonyOS. This suggests that effective mobile application migration requires platform-aware reconstruction and validation, rather than relying solely on LLM-based code generation. 





\balance
\bibliographystyle{ieeetr}
\bibliography{sample-acmsmall-conf}


\end{document}